\documentstyle[fleqn,psfig]{tp}

\def\bi{\begin{itemize}}
\def\ei{\end{itemize}}

\def\thedemobiblio#1{\smallskip\par
 \list{}{\labelwidth 0pt \leftmargin 1em \itemindent -1em \itemsep 1pt}
 \small \parindent 0pt
 \parskip 1.5pt plus .1pt\relax
 \def\newblock{\hskip .11em plus .33em minus .07em}
 \sloppy\clubpenalty4000\widowpenalty4000
 \sfcode`\.=1000\relax}

\let\ssec=\subsection

\newcount\japif
\def\japeqn{\ifnum\japif=1
\begin{equation}\global\japif=0 \else
\end{equation}\global\japif=1\fi}

\def\japref{\item}
\def\gs{\mathrel{\lower0.6ex\hbox{$\buildrel {\textstyle >}
 \over {\scriptstyle \sim}$}}}
\def\ls{\mathrel{\lower0.6ex\hbox{$\buildrel {\textstyle <}
 \over {\scriptstyle \sim}$}}}
\newcount\equationo
\def\m@th{\mathsurround=0pt }
\def\eqalign#1{\null\,\vcenter{\openup1\jot \m@th
 \ialign{\strut\hfil$\displaystyle{##}$&$\displaystyle{{}##}$\hfil
 \crcr#1\crcr}}\,}

\def\apj{ApJ}

\def\mn{MNRAS}

\newcommand{\kms}{\,km~s$^{-1}$}      

\newcommand{\sun}{\mbox{$\odot$}}
\newcommand{\et}{{\it et al.}~}

\def\ltsima{$\; \buildrel < \over \sim \;$}
\def\simlt{\lower.5ex\hbox{\ltsima}}
\def\gtsima{$\; \buildrel > \over \sim \;$}
\def\simgt{\lower.5ex\hbox{\gtsima}}
\def\arcs{$''~$}
\def\arcm{$'~$}

\begin{document}

\twocolumn[
\title{Observations of Galaxy Clustering at High Redshift}
\author{C. C. Steidel \\
{\it Palomar Observatory, California Institute of Technology} \\
{\it Caltech 105-24, Pasadena, CA 91125, USA}
}

\vspace*{16pt}   

ABSTRACT.\
This paper reviews the current status of measurements of galaxy clustering
at high redshifts ($z \simgt 0.3$). The focus is on the inherent limitations
in the observation and interpretation of the ``evolution of clustering''.
It is likely that results from the first attempts to characterize galaxy
clustering beyond the ``local'' universe 
have been significantly limited by sample variance, as
the difficulty in assembling large samples over large volumes is exacerbated
as the observations become more challenging.  
It is also argued that, because of the complicated relationship between
galaxies and mass (i.e., bias), and the surprising degeneracies among different
popular cosmological models, 
it is likely that studies of galaxy clustering as a function
of cosmic epoch will never be useful for strong discrimination between
different cosmological models. On the other hand, observations of galaxy
clustering are capable of testing basic ideas about how (and where) galaxies
form. Galaxy formation, as opposed to cosmography,  
will probably remain a fundamental question even beyond the MAP and Planck era.  
\endabstract]

\markboth{C. C. Steidel}{Clustering at High Redshift}

\small

\section{Introduction}

We are clearly living in an era in which the relatively nearby universe
will be mapped out with exquisite precision with new dedicated telescopes
and instruments. The aims of such surveys seem clear: to use galaxies
as a means to map out the large--scale structure of the universe, in the hope
that in so doing one can understand the details of the relationship between
observable galaxies and the overall matter distribution, and ultimately, to
test theories of structure formation and to measure the details of the
power spectrum of mass fluctuations on scales that are capable of testing theoretical
ideas about the origins of the fluctuations.  Even at ``zero'' redshift, as
we have heard as a major theme at this conference, 
there is considerable argument about the degree
to which galaxies should be trusted as tracers of mass; in particular, 
we now know that the clustering properties of galaxies are not universal,
but depend on galaxy color, luminosity, and other messy astrophysical properties
that are correlated with, but not direct proxies for, mass. 
The days of treating all galaxies in a redshift survey as identical test particles
are almost certainly over; the degree to which one must worry about things like population
mixes and luminosity and color segregation depend on the nature and scale of the
measurement being made. The point here is that to understand large--scale structure
as traced by galaxies, one needs also to understand something about the galaxies
themselves. This is either annoying, or incredibly interesting, depending on one's perspective.  

For redshifts where the evolving properties of galaxies begin to be important,
and where the quantity of available information drops off precipitously, 
we are still in very much an exploratory phase. At $z \simgt 0.3$, there has
been a great deal of progress, but we are happy
enough to have any measurements at all-- the kind of careful scrutiny to which
the current and future major ``local'' redshift surveys have been subjected has not yet
descended. This relatively information--starved regime is the subject of this
short review.  

Now that observations of large samples of high redshift galaxies are feasible from
a purely technical standpoint, it is worth revisiting the question
of just what one learns by studying the evolution of clustering with
redshift. It is also worth considering in some detail how the selection
of galaxies, and the overall design of a survey, may significantly influence results. 
The organization of this review is to first discuss the status of observations
of galaxy clustering at $z \simlt 1$; the observations here consist of
traditional apparent--magnitude selected redshift surveys designed primarily
for studying galaxy evolution, to new wide--angle imaging surveys and
applications of photometric redshift techniques. This will be followed by a discussion of results
and prospects using large photometric and spectroscopic surveys
at $z \simgt 3$. 

\section{``Evolution'' of the Correlation Function}

Because of the relatively small numbers of galaxies in most of the high redshift
samples, simple statistics are generally used to describe the overall level
of clustering. Many have described the clustering in terms of the Groth \& Peebles (1977)
parameterization of the two--point correlation function,
\begin{equation}
\xi(\rm{r,z}) = \left( {\rm{r}} \over {\rm r_o(z)} \right)^{\gamma} (1+{\rm z})^{-(3+\gamma+\epsilon)}
\end{equation}
where $r$ is the co-moving coordinate distance, $r_0$ is the co--moving correlation length,
$\gamma$ is the slope of the correlation function and $\epsilon$ is the ``evolutionary parameter''.
In the context of this parameterization, with $\gamma=-1.8$, one obtains the usual limiting cases
of $\epsilon=-1.2$ for clustering fixed in co--moving coordinates (i.e., ${\rm r_0(z)}=$constant), 
$\epsilon=0$ for clustering fixed in physical (proper) coordinates (${\rm r_0 \propto (1+z)^{-2/3}}$),
and $\epsilon=0.8$ for linear growth of clustering in an Einstein--de Sitter universe [${\rm r_0(z)
\propto (1+z)^{-1}}$, approximately]. This would be a very good way of thinking about the
evolution of clustering if it were the case that 1) one were seeing the same galaxies at all
redshifts ${\rm z}$, and 2) if galaxy clustering were a monotonic function of scale factor.
In practice, 1) is almost certainly not the case, since selection effects and galaxy evolution
conspire to bring different types (masses?) of galaxies into and out of samples as a function
of redshift, and quite probably 2) is also not satisfied. 

As has been discussed by several
others at this conference (e.g., Peacock in these proceedings), 
a generic result of simulations and/or analytic 
models is that the correlation function for dark matter halos actually evolves very
differently from the correlation function of the mass, and actually passes through a minimum
at intermediate redshifts after having begun in a highly biased state when fluctuations of
the particular mass threshold were very rare (see, e.g., Brainerd \& Villumsen 1994; Bagla 1998b). 
In this picture, the evolution
of the bias of the dark matter halos out-paces the growth of matter fluctuations (so that
the galaxy clustering becomes {\it weaker} in co--moving units with decreasing redshift) 
until a characteristic
redshift at which the mass threshold is a $\sim 1\sigma$ fluctuation, after which the clustering
strength increases again. For halos typical of bright galaxies today, $\sim 10^{12}$ M$_{\sun}$,
this clustering minimum is expected to occur near $z \sim 1$. Thus, under this picture, the
{\it observed} evolution of clustering would depend upon the characteristic halo mass (or mix
of mass scales)
traced by galaxies satisfying the particular sample selection criteria at each redshift. Even if one
could somehow isolate galaxies of fixed mass, the evolution of the clustering would not
fit in well with the $\epsilon$ parameterization shown above. In the real world, one is likely
to be observing a complicated mix of galaxies/mass scales that is likely to be changing as
a function of redshift, so that what one observes in a sample is a superposition of such 
complicated evolutionary sequences. The prediction would then be that describing the evolution
of clustering with a single value of $\epsilon$ that results from a ``best--fit'' to a
heterogeneous sample (whether it be heterogeneous with respect to color, luminosity, etc.)
is not likely to provide information that is particularly useful in understanding what is going
on.  Many have now suggested dispensing with the $\epsilon$ parameterization altogether; this
is an excellent suggestion. 

\section{Clustering at $z \simlt 1$}

Until very recently, the only information on the clustering of distant
galaxies that was available was based on studies of the angular
correlation function of faint galaxies (e.g., Efstathiou \et 1991, Efstathiou 1995, 
Brainerd \et 1995, Postman \et 1998). The general result
was much weaker clustering amplitude for faint galaxies than seen 
in local galaxy surveys, but because of the unknown extent of projection
effects due to a lack of detailed knowledge of the redshift
distribution $N(z)$ [in some cases the clustering results were used to place
constraints on the behavior of ${\rm N(z)}$] and the uncertainty that the galaxies seen at faint
magnitudes are the same objects counted in local galaxy surveys, the
implications were somewhat ambiguous. A tendency was seen in several of the above
surveys for a flattening in the angular correlation function amplitude at the
faintest magnitudes; however, the degeneracy between the issue of the
redshift distribution and the evolution of the clustering does not permit
the solution for one without knowledge of the other. This degeneracy/projection
problem in the imaging surveys can be overcome, to some extent, by using multiple
colors to assign photometric redshifts, as discussed by A. Connolly at this conference.  
While there are some limitations to the photometric redshift technique, it appears
to work very well for $z < 1$ (e.g., Hogg \et 1998 and references therein). 

\begin{table*}
\caption[]{Non-exhaustive summary of recent galaxy clustering results to $z\sim 1$. Correlation
lengths are in co--moving units.}
\centering
\begin{tabular}{lcrlc}
\hline
Survey & $\langle z \rangle$ & ${\rm N_{gal}}$ & ${\rm r_0/h}$ Mpc & Comments \\
\hline 
\hline
CNOC-1 & 0.35 & 140 & $2.7\pm0.6$ & 1 \\
CFRS  & 0.34 & 186 & $1.8\pm0.2$ & 2 \\
                           & 0.62 & 196 & $1.8\pm0.2$ & 2  \\
                           & 0.86 & 130 & $1.9\pm0.2$ & 2  \\
\,\,(Full CFRS Sample)    & 0.53 & 591 & $2.2\pm0.1$ & 2 \\
Hawaii Deep    & 0.34 &  &  $3.9\pm0.2$ & 3 \\
                & 0.62 &  & $3.2\pm1.1$ & 3 \\
                & 0.97 & & $2.8\pm1.2$ & 3 \\
		& 1.39 & & $2.4\pm1.2$ & 3 \\
\,\,Red Sub-sample & 0.6 & $\sim 100$ & 3.8 & 3 \\
\,\,Blue Sub-Sample & 0.6 & $\sim 150$ & 1.4 & 3 \\
KPNO I-band    & (0.5) & 4e5 & $4.5\pm0.6$ & 4 \\ 
CNOC-2 (bright) & 0.35  & $\sim 1500$ & $5.0\pm0.2$ & 5 \\
CNOC-2 (faint)  & 0.35  & $\sim 1500$ & $3.6\pm0.2$ & 5 \\
\hline
\end{tabular}
\begin{tabular}{l}
1) Shepherd \et (1996); found best--fit $\epsilon \sim 1\pm1$ \\
2) Le F\`evre \et (1996); 5 10\arcm\ fields, to I$_{\rm AB}=22.5$.\\ 
\,\,Found $\gamma=-1.6$,
$\epsilon \sim 0-2$, and no color segregation for $z > 0.3$.\\ 
3) Carlberg \et (1997); K-selected sample, total $\sim 250$ z's. \\
\,\,Assumes fixed $\gamma$, $q_0=0.1$. Clear color segregation. \\
4) Postman \et (1998); $w(\theta)$ de--projection using CFRS \\ 
\,\,redshift distribution and magnitude cuts. \\
5) Carlberg \et (1998); sample split into ``bright and faint''\\
\,\,at ${\rm M_R}=-20$; find best fit evolutionary model is\\
\,\,${\rm r_0(z) \propto (1+z)^{-0.3\pm0.2}}$.\\
\end{tabular}
\end{table*}

A first attempt at using a deep spectroscopic survey for measuring the spatial
correlation function of faint galaxies (in this case at a median redshift
of ${\rm z =0.16}$) was made by Cole \et (1994), 
who obtained a correlation function that was indistinguishable, in co-moving
units, from the local correlation function. There have been a number of subsequent
spectroscopic surveys which have addressed galaxy clustering at $z \simgt 0.3$, many
of which are summarized in Table 1. This summary includes minimal details describing
the results on the co--moving correlation lengths, in particular, from the various
surveys. There are other large spectroscopic surveys which reach similar
redshift depths whose results were in preparation at the time of this meeting
(e.g., Cohen \et 1999, Small, Sargent, \& Ma 1999) and which have not
yet reported actual measured correlation functions in the literature. 
However, the Caltech
Deep Redshift survey (a K-selected sample) of Cohen \et (1999) has reported a strong tendency for objects
that are red in their optical/IR colors and have absorption-dominated spectra
to preferentially inhabit the most prominent structures in redshift space; the
kinematics of these structures suggest that they are groups or poor clusters.
There is clear evidence for both luminosity and color segregation in the clustering
properties, but the quantitative comparison with the surveys presented in Table 1
is not yet possible. 

A benchmark piece of work in terms of measuring the actual evolution of the correlation
function as a function of redshift, within the same survey, was that of Le F\`evre \et (1996)
from the Canada-France Redshift Survey (CFRS). As can be seen in Table 1, the CFRS team
had a large enough sample, over a large enough redshift range, that the data could be binned
into redshift subsets. The measurements of the correlation lengths versus redshift, and
of the sample as a whole, seemed to show strong evolution of the galaxy
correlation function in the sense that the correlation strength was significantly weaker
in the past; it was argued that is was very unlikely that they were seeing different
galaxies as a function of redshift within the sample, so that for the first time
one could see the growth of clustering of galaxies directly. Interestingly, the CFRS
saw no evidence for color segregation (based on optical colors) of the clustering of galaxies for redshifts
beyond $ z\sim 0.3$. 

However, a glance at Table 1 makes one worry slightly. Compare, for example, the
preliminary CNOC-2 results at $z \sim 0.35$ with the lowest redshift bin
of the CFRS survey, or with the results of the Hawaii Deep Survey K--band sample.
The CNOC-2 sample finds significantly stronger clustering even for the ``faint'' 
sub--sample, and much stronger clustering for the ``bright'' sub--sample; the 
Hawaii Deep Survey, albeit a relatively small sample, finds clear evidence for
different clustering for red and blue sub--samples, and only the blue sub--sample
exhibits a correlation length comparable to the CFRS results at comparable redshifts.  
If one takes all of the results together, there seems
to be relatively weak, but significant, evolution of clustering
in the sense that it is slightly less strong in the past. But individual survey
results exhibit a very wide scatter--not at all consistent with
the quoted uncertainties on the measurements, and inferred values of the evolutionary
parameter $\epsilon$ are all over the map, from slightly negative to strongly
positive. 
In the largest spectroscopic sample, CNOC-2, there appears
to be both luminosity and color segregation present, in the sense that redder and more luminous
galaxies are more strongly clustered (as is seen in some local galaxy surveys, e.g. Loveday \et 1995).   
This survey does not reach high enough redshifts to make detailed comparisons at the median
redshift of the CFRS sample.

If there are lessons learned from the work that has been completed recently
in the $z \simlt 1$ regime, it is that most of the samples have probably not
been large enough to yield universal results on the galaxy correlation function--
there is a trend of increasing measured ${\rm r_0}$ at a given redshift
with increasing sample size, a telltale sign that sample variance has been
a problem. The results of the very large photometric survey by Postman \et (1998)
perhaps illustrates the problem best, in that one can isolate many independent sub-samples
that are as large as (for example) the CFRS, and it is clear that in a sample the
size of the CFRS there is a quite large probability of measuring clustering strength
that is significantly smaller than that observed over very large volumes. 
Using the CFRS photometric selection criteria for their angular correlation function,
and the CFRS--observed N(z) for de--projecting to form the real--space correlation
function, Postman \et obtain
a value for the correlation length that is more than two times larger, at the
median redshift, than that obtained by Le F\`evre \et, a difference that is
significant at about the 4$\sigma$ level if one takes the error bars at
face value. On the other 
hand, the same large imaging survey of Postman \et, while making a measurement of unprecedented precision
in terms of sheer reduction in sample variance and counting statistics,
is unable to distinguish between a number of widely different models for
the growth of clustering; this indicates the inadequacy of the $\epsilon$
parameterization in general, and the need for cutting down on projection effects
that hamper deep imaging surveys intended for clustering analysis. One very
sensible means of overcoming this limitation is to use multi-bandpass imaging and
the photometric redshift method to isolate cosmic epochs, galaxy luminosity classes,
and color cuts, exploring the behavior of the clustering as a function of
galaxy properties in a multi-dimensional manner. In doing this, one clearly
increases the noise in any clustering estimate, but it is almost certainly
worth sacrificing {\it precision} in favor of {\it information content}. 
Most useful of all, although rather painstaking in terms of the resources necessary
to obtain the observations, are wide angle, deep spectroscopic surveys such
as CNOC--2 
(and future surveys such as that planned by the DEEP collaboration--see Davis \& Faber 1998)  
where in addition to simple correlation statistics, precision galaxy redshifts might yield more detailed dynamical information that might
help relate the galaxy clustering to the clustering of the dominant dark matter component on
small scales. Both large volumes {\it and} accurate redshifts, with a good sampling of redshifts
using sensible selection criteria, will be invaluable.

\section{Clustering at $z >> 1$}

Given that we do not really understand the details of what is going on with
galaxy clustering at intermediate redshifts, why would one want to bother
exploring the higher redshift universe, where the problem of making
clear evolutionary connections to galaxies at the present
is even more difficult?  The simplest answer is that the pure novelty
of compiling a large sample of high redshift galaxies would be bound
to yield something interesting; naively, one might have thought that 
one of the cleanest possible tests of the idea that matter fluctuations
grow by gravitational instability would be to obtain a ``snapshot'' of
galaxy clustering at very high redshifts, where one might expect
the clustering to be much weaker if gravity really were the dominant
factor in producing the structure observed in the local universe. 
Also naively, one might have expected that the amount by which the
clustering should be different would be rather sensitive to 
$\Omega_m$, with smaller differences expected for lower $\Omega_m$. 
All of this would be true if galaxies formed at infinite redshift and
then evolved quiescently to the present epoch, acting like
conserved test particles. As it turns out, the clustering {\it is}
a sensitive test of our collective wisdom about how galaxies form,
but probably a relatively poor cosmological discriminant. 

The method used for compiling samples of very high redshift galaxies to date has been
almost exclusively the ``Lyman--break'' technique, where
one makes use of the essentially guaranteed ``break'' in the spectrum
of high redshift star forming objects at 912 \AA\ in the rest frame
due to photo-electric absorption both in the galaxy itself and in the intergalactic
medium. The feature is so strong that the coarse spectrophotometry
allowed by broad--band imaging can be used to isolate particular
ranges of redshift that can be controlled based on the adopted filter system.
The technique as successfully implemented 
has been described recently in many places (e.g., Steidel, Pettini, \& Hamilton 1995, Steidel \et 1996,
1998b, Madau \et 1996),
so I will not go into any details here. Basically, the method is a highly efficient means
of selecting a nearly volume--limited sample of objects on the basis of their rest--frame far--UV
luminosity. As with any other galaxy sample, it is important to understand what selection
effects are implicit in the detection technique: here, one is quite insensitive to 
stellar mass, but sensitive almost exclusively to unobscured high--mass star formation.   
Extremely dusty galaxies, or galaxies which have ceased forming stars prior to the epoch
at which they are observed (or those going through a quiescent phase between star formation
episodes) are unlikely to be included. 

The particular implementation of the Lyman break technique for which the most
data have been obtained to date selects
galaxies in the redshift range $2.7 \simlt z \simlt 3.4$, as shown in Figure 1. 
\begin{figure*}
\centering\mbox{\psfig{file=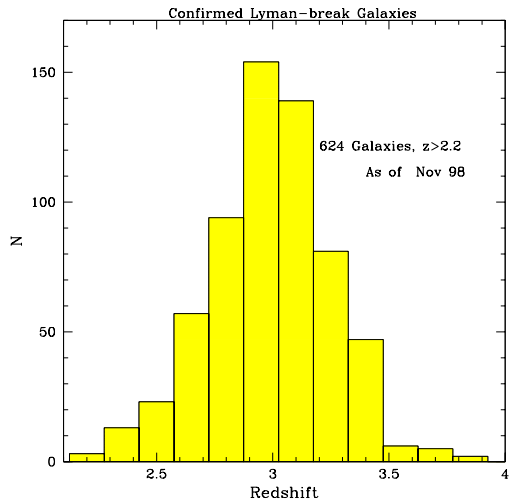,height=7cm}}
\caption[]{Redshift histogram of Lyman break galaxies in the $z \sim 3$ sample,
selected using color criteria in the $U_nG{\cal R}$ color--color plane
from wide--field ground--based images. All of the confirming spectroscopic redshifts
were obtained with the Keck telescopes and the Low Resolution Imaging Spectrograph
(Oke \et 1995).}
\end{figure*}
The primary goal of the survey is to accumulate a large enough sample of high redshift
galaxies that proper statistics on the luminosity distribution, spectral properties,
reddening, and, most relevant here, their large--scale distribution, are possible. 
The survey is
in many ways similar to the CFRS in design: there are 5 primary survey regions, 
typically
9\arcm\ by 18 \arcm\ in angular size--- comparable to that of the CFRS, but the actual transverse
{\it co--moving} scale is much larger on account of the much higher redshift of the survey.
For example, the co--moving size of a CFRS field at $z \sim 0.6$ is $4.5{\rm h}^{-1}$ Mpc on
a side (inside which the CFRS densely sampled 3 10\arcm\ by 2\arcm\ strips), 
as compared to $11.6{\rm h}^{-1}$ Mpc by $23.3{\rm h}^{-1}$ Mpc for the LBG fields
at $z \sim 3$, for $\Omega_m=0.3$, $\Omega_{\Lambda}=0.7$. The CFRS depth along
the line of sight, on the other hand, is between 2 and 4 times larger than in the LBG
fields, depending on cosmology. Practical
matters, mainly having to do with the faint apparent magnitudes of even the more
luminous LBGs,  limit the sampling density for galaxies at $z \sim 3$ in these larger volumes, so
that scales much smaller than the field size are not probed well. This bears significantly
on what types of clustering statistics can be measured well using the data.   
Each LBG survey field samples an effective co--moving volume of $\sim 2.2 \times 10^{4}$ h$^{-3}$ Mpc$^{3}$
for an Einstein-De Sitter model ($\sim 8.3\times 10^{4}$ h$^{-3}$ Mpc$^{3}$ for $\Omega_m=0.3$
and $\Omega_{\Lambda} = 0.7$), so that the total volume surveyed is somewhere between
$10^5$ and $10^6$ h$^{-3}$ Mpc$^{3}$.   

Within these survey regions, to the adopted apparent magnitude cutoff of ${\cal R}_{\rm AB}=25.5$,
there are $\sim 1500$ photometrically selected candidates, and the aim is for a total spectroscopic
sample of $\sim 750$. Although the survey is not yet completed, several interim results have 
already been published. Many of the results have been discussed elsewhere, e.g. Steidel \et 1998a,b,
Adelberger \et 1998, Giavalisco \et 1998, so the discussion here will be brief.  

\begin{figure*}
\centering\mbox{\psfig{file=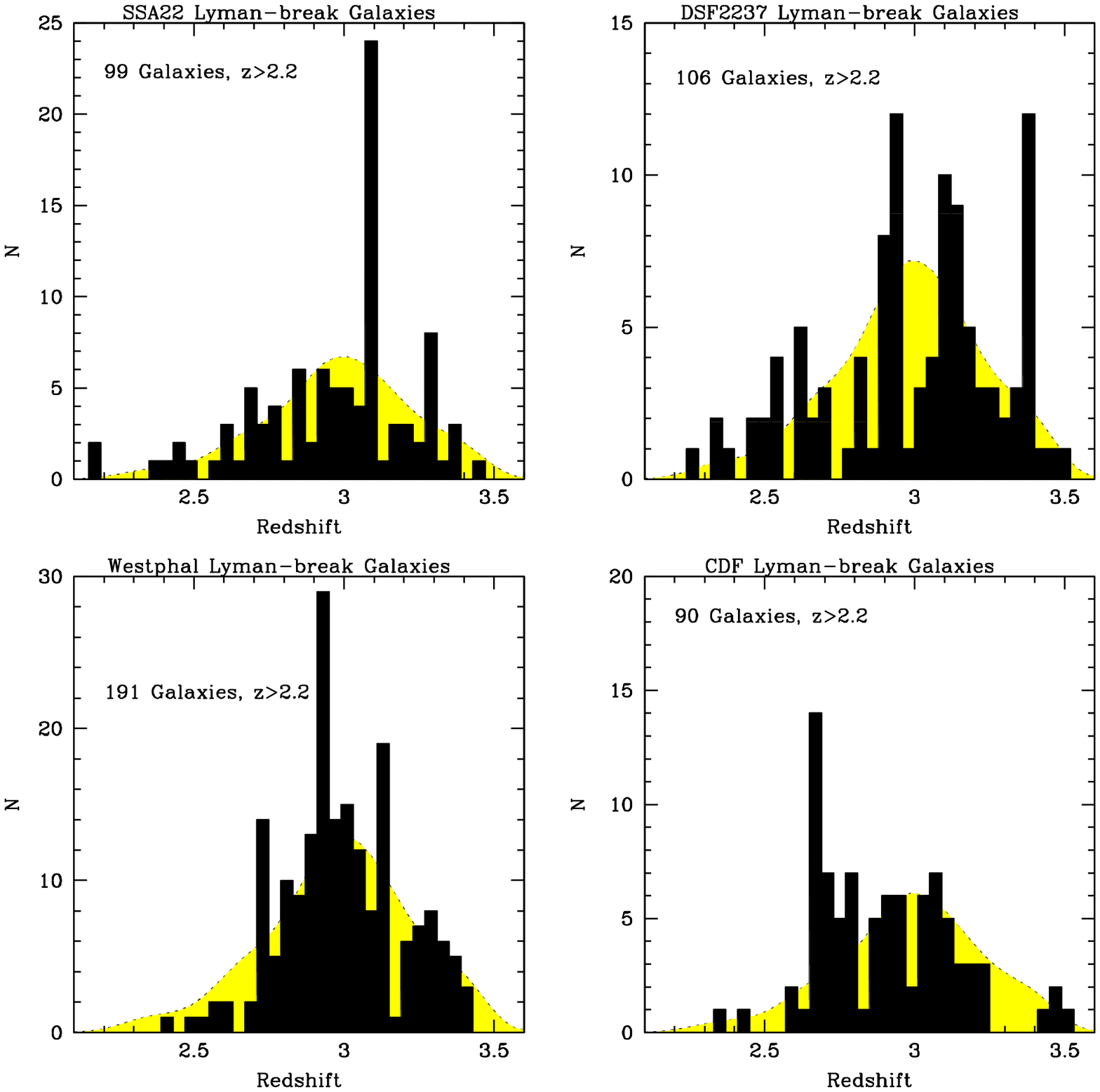,height=8cm}}
\caption[]{Redshift histograms in individual fields of the $z \sim 3$
Lyman break galaxy survey. Each field is 150-225 square arc minutes in
size. The light smooth histograms in each case is the overall redshift
selection function for the survey, normalized to contain the same number
of galaxies as observed in each field. Note the presence of strong
``spikes'' in the redshift distribution, and a few significant
``voids'' as well. Each bin encompasses a volume
on the order of 1000${\rm h}^{-3}$ Mpc$^3$}
\end{figure*}

First, as can be seen in Figure 2, the Lyman break galaxies are strongly clustered, and
large over--densities, or ``spikes'' in the redshift
distribution are evident in each survey field. The galaxies within these over-densities
are not obviously concentrated on the plane of the sky, and angular correlations
of the photometric samples of Lyman break galaxies have rather poor S/N because of
the aforementioned low surface density and therefore poor sampling of the small
scales where most of the angular correlation signal would lie. At present, the
most robust statistic for evaluating the level of clustering for the LBGs is
a ``counts--in--cells'' analysis. Here one simply counts the number of
objects in cubical cells of roughly $10{\rm h}^{-1}$ Mpc on a side (the scale being defined
by the transverse size of the survey field, which of course varies as a function of
cosmology) in the spectroscopic sample, corrects for the shot-noise contribution
to the variance and modest redshift--space distortions, 
and evaluates $\sigma_{\rm cell}$ (see Adelberger \et 1998). The variance
in galaxy counts (relative to the overall expectation value from the selection function)
in cells of this size is very closely related to the commonly--used ``$\sigma_8$''
statistic for normalizing mass fluctuations using cluster abundances (see, e.g., White \et 1993,
Eke, Cole, \& Frenk 1996). Since it is straightforward to compute the expected mass
fluctuations on the same scales at $z \sim 3$ (using present--day cluster
normalization) for a given cosmology, one can essentially ``read off'' the
required galaxy effective bias on the scale of a ``cell'' from a plot similar to the
ones shown in Figure 2. The most recent numbers resulting from such an analysis
of the current LBG survey sample, assuming cluster normalization of Eke, Cole, \& Frenk (1996) are
\begin{equation}
 {\rm b_{eff}} = \left\{\begin{array}{rl}
    5.2\pm 0.9 & \hbox{$\Omega_m=1$} \\
    3.6\pm 0.6 & \hbox{$\Omega_m=0.3$,\,$\Omega_{\Lambda}=0.7$} \\
    1.7\pm 0.4 & \hbox{$\Omega_m=0.2$} \\
 \end{array}\right.\;.
\end{equation}
where ${\rm b_{eff}}$ is the effective linear bias on scales of $\sim 10{\rm h}^{-1}$ Mpc. 

Thus, it can be seen that bright LBGs {\it must} be strongly biased tracers of mass
fluctuations in order to be accomodated easily by standard hierarchical models.
It is somewhat more difficult, given the nature of the data, to turn this into
a correlation function (although, of course, ${\rm b_{eff}}$ is equivalent to an
integral over the correlation function over the cell volume)--while we have attempted
to de-project the angular correlation function of LBGs to obtain a real--space
correlation function (Giavalisco \et 1998), the current sample is not well--suited to
measuring $w(\theta)$ nor $w_p(\theta)$ accurately. The central problem is that the depth
of one of our survey fields (in co--moving units) far exceeds its width, and as a result
the vast majority of angular pairs are simply chance projections of galaxies at
very different redshifts. For example, even at separations as small as 20\arcs\, 
approximately 90\% of galaxy pairs in our sample are chance projections
(Giavalisco \et 1998). Including these chance pairs in a clustering analysis--- as is
required for ${\rm w(\theta)}$ and some forms of ${\rm w_p}$ --- results in a disastrous
reduction in signal--to--noise ratio. This seems an unnecessarily high price for
avoiding peculiar velocity distortions, which are after all relatively
minor on large scales at these redshifts. Measuring ${\rm w(\theta)}$ is a reasonable
approach when the sample is large enough that random fluctuations in the number
of chance pairs are small compared to the number of true pairs, and is
often the only approach when only a small fraction of the galaxies in the sample
have redshifts; however, neither condition is true for the $z \sim 3$ LBG sample. 

Thus far we have been quoting a number for the correlation lengths that is based
on assuming a value for the slope $\gamma$ of the correlation function (Adelberger \et 1998),
and these number range from 4--6h$^{-1}$ Mpc, with the lower end of the range applying to
Einstein-de Sitter and the upper range to low $\Omega_m$ models. We plan to do a more
careful job with this when the spectroscopic sample has been completed, which we anticipate
will be  quite soon.
Regardless of the precise values for $r_0$, 
the clustering of the $z \sim 3$ LBGs is as strong, or stronger, than most
local galaxy samples, and significantly stronger than most of the intermediate redshift
numbers. It is still unclear whether we have sampled enough volume to asymptote
to the ``truth'', but these correlation lengths are likely
to be firm lower limits. 

Strong clustering/bias is expected theoretically for rare peaks in the density field (Kaiser 1984),
and many theoretical papers, both predating and interpreting the clustering observations, can easily
explain the strong clustering of the LBGs through this ``high peaks'' biasing (e.g.,
Baugh \et 1998, Coles \et 1998, Wechsler \et 1998, Bagla 1998a, Mo and Fukugita 1996, Jing
\& Suto 1998, Governato \et 1998, Mo, Mao, \& White 1998, Katz, Hernquist, \& Weinberg 1998). 
One can get quite good agreement with both the abundance and clustering properties
of dark matter halos (using either N-body or analytic techniques) and the real galaxies provided that the typical LBG in the sample is associated with a dark matter halo mass scale of
$\sim 10^{12}$ M$_{\sun}$ (Steidel \et 1998a,b; Adelberger \et 1998; Mo, Mao, \& White 1998). 
This good agreement suggests that there should be a monotonic relationship between
dark matter halo mass and UV luminosity, and that most, if not all, dark matter halos of
a given mass contain a LBG exhibiting a star formation rate with relatively small scatter
(Adelberger \et 1998). This provides empirical evidence that the use of star formation
prescriptions that are based on the dark matter halo properties (as in most semi-analytic
models of galaxy formation) may be on the right track.

A power spectrum
shape parameter of $\Gamma \sim 0.2$ is most consistent in matching the inferred
bias and the abundance of galaxies and dark matter halos, but otherwise surprisingly little
difference is expected for the clustering of objects of a given abundance 
among the currently popular dark matter models which have
this kind of shape parameter (i.e., $\tau$CDM, open CDM, $\Lambda$ CDM). 
As discussed by 
Steidel \et 1998b, Adelberger \et 1998,  and Giavalisco \et 1999, 
a more stringent test of such a simple
association of LBGs with dark matter halos in a hierarchical model would come from
examining the clustering of much fainter LBGs, which would presumably trace much 
smaller mass dark matter halos which should be significantly less clustered at high
redshift. Preliminary indications show that most of the models remain consistent with
the data when faint LBG samples
from the HDF are compared with the ground based results, with significantly smaller
correlation lengths for objects with abundances $\sim 20$ times larger than the bright
galaxies in the ground--based survey. Larger samples, particularly
of the faint objects, will be required in order to be able to exert much pressure on
any of the currently popular dark matter models. 

On the other hand, there is a very substantial difference among the various models 
for the {\it masses} of objects
of a given abundance and clustering level.
This difference is large enough (a difference of a factor of $\sim 3$ in circular
velocity) between low $\Omega_m$ models and $\tau$-CDM that observations of line
widths (even with all of the inherent uncertainties in using them for
dynamical mass estimates) may be able to resolve the degeneracy. The main problem
is that line widths are essentially always providing lower limits on $v_c$, and
some theoretical predictions suggest (e.g., Mo, Mao, \& White 1998) that the {\it observed} 
line widths may not be radically different despite the very different $v_c$
because of the fact that the highest star formation efficiency would occupy
regions that are still on the rising part of the rotation curve.   Some observations along
these lines have already
been attempted using the familiar nebular lines in the rest--frame optical (observed
near-IR) (Pettini \et 1998), but the advent of IR spectrographs on 8--10m
telescopes should result a huge amount of progress in this area.  

The UV spectra of LBGs, on the other hand, represent possibly the most frustrating limitation
of the spectroscopic samples. While in principle the redshift accuracy achievable at
$z \sim 3$ is the same as one could achieve at intermediate redshift, the problem is
that essentially none of the commonly--observed far--UV lines is trustworthy
as an indication of the systemic redshift of the galaxy. We have estimated the intrinsic
uncertainty (independent of measurement errors) to be on the order of $\sim 300$ \kms.
This means that it may be difficult to explore any statistics based on small-scale
dynamics (e.g., pairwise velocity dispersions) without wholesale IR spectroscopy.

\ssec{General Implications}

The nature and clustering of the $z \sim 3$ LBGs
are very consistent with the overall ``paradigm'' that galaxies would form at the highest, ``biased''
peaks in the dark matter distribution at early epochs, and that these objects should be strongly
clustered at high redshift. These clustering properties, together with the
observed space densities, imply the individual galaxies are associated
with dark matter halos of the order of $\sim 10^{12}$ M$_{\sun}$. Within the context
of these models, a large fraction of the LBGs seen in the bright ground--based
samples would end up in richest environments in the present--day universe, and
the prominent ``spikes'' at $z \sim 3$ are likely to be the progenitors of
present--day rich galaxy clusters (e.g., Steidel \et 1998a, Governato \et 1998). The incidence
of these prominent over-densities is indeed broadly consistent with this hypothesis. 
Observations of the clustering properties as a function of space density for LBGs over
a wide range of luminosities have the potential to measure the shape of the power spectrum
on scales of $\sim 1-10$h$^{-1}$ Mpc (i.e., between galaxy and cluster scales) if the case
can be made that the UV luminosity really is a good proxy for dark matter halo mass. 
Here again the problem is somewhat circular, in the sense that the dark matter models
can be tested rigorously only if some observable property of the galaxies can be closely
tied to the mass, but if one knew the underlying structure of the dark matter, it would
be possible to close in on how star formation is related to the dark matter distribution.
At present, the assumption that UV luminosity and dark matter halo mass are very closely
related, with a power spectrum shape very close to that which works best to explain
the local large scale structure (see e.g., Peacock \& Dodds 1996), works very well indeed,
but this solution cannot be said to be unique at this point in time. 

\begin{figure*}
\centering\mbox{\psfig{file=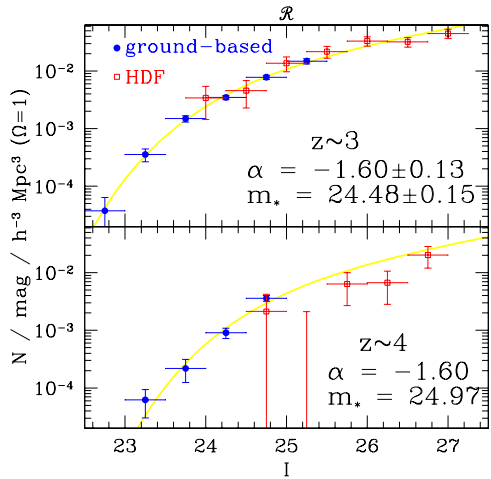,height=8cm}}
\caption[]{Luminosity functions obtained by combining our ground--based, wide--area
surveys, with data on faint objects in the Hubble Deep Field. The HDF points
are based on the catalogs of LBGs presented by Dickinson (1998) and Madau \et 1998,
but we have reanalyzed the effective survey volumes with knowledge of the
true color distributions of the LBGs based on our spectroscopic samples. The
bright ends of the $z \sim 3$ and $z\sim4$ luminosity functions are strikingly
similar, in both shape and normalization (the $z \sim 4$ curve is simply
the fit at $z \sim 3$ shifted by the distance modulus between $z=3.04$ and
and $z=4.13$, with the normalization multiplied by 0.8). 
The integrals of UV luminosity to the equivalent of $I_{AB}=25$ in
the higher z sample is within about 20\% of that over the same luminosity
range at $z \sim 3$, independent of cosmology. Note the indications that the HDF
is underdense in $z \sim 4$ galaxies.   
}
\end{figure*}
\begin{figure*}
\centering\mbox{\psfig{file=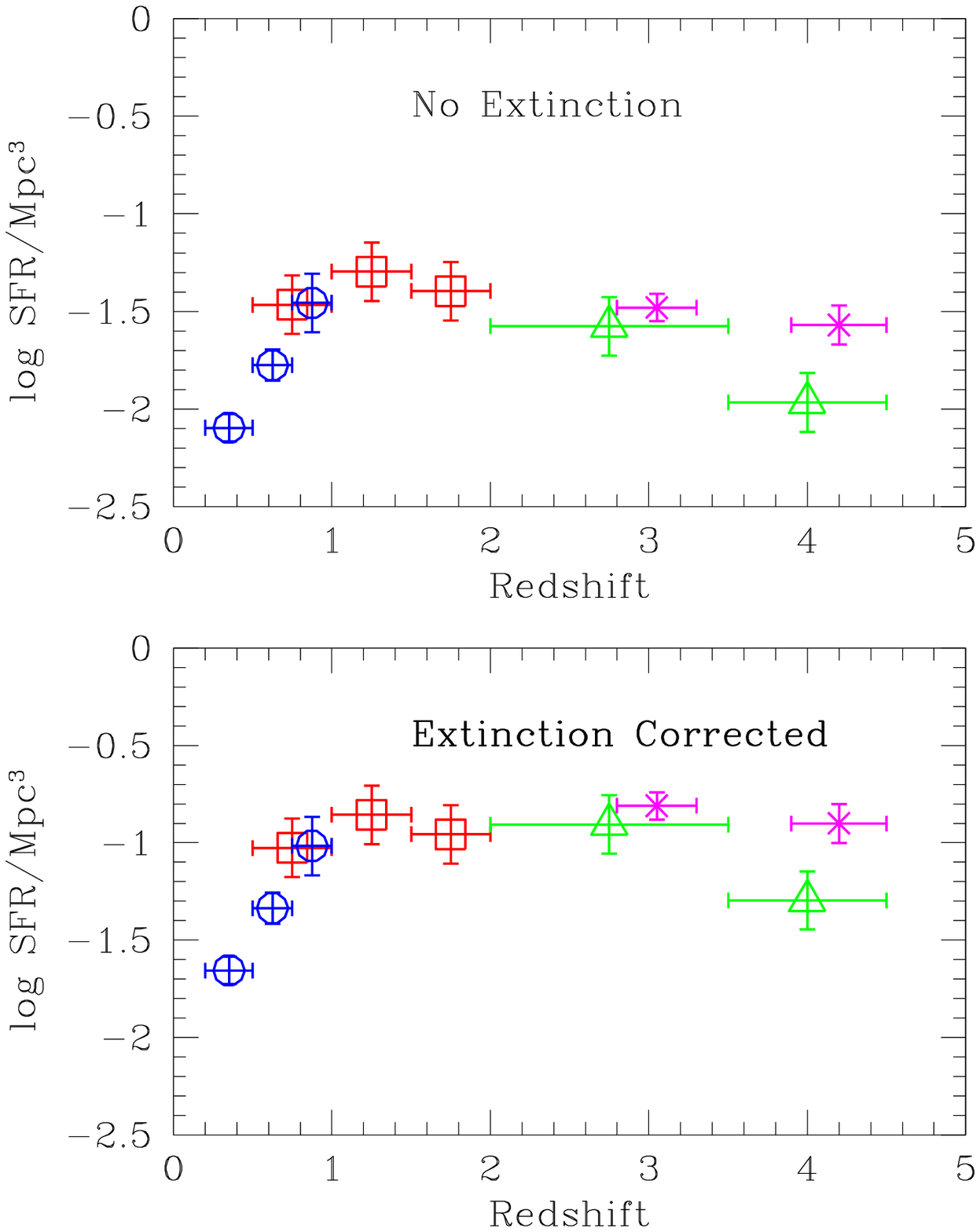,height=8cm}}
\caption[]{A revised version of the ``star formation history'' diagram, with
new points from the large ground--based surveys indicated with the crosses.
The circles come from Lilly \et (1996), the squares from Connolly \et (1997),
and the triangles from Madau \et (1998). Note that with internally consistent
corrections for extinction (see Steidel \et 1999 for details), there is
no indication for any significant change in the universal star formation density for
any $z >1$.
}
\end{figure*}

\section{$z \sim 4$ and Beyond}

At the time of this writing, a handful of galaxies have been identified
beyond $z \sim 5$. Should one be thinking about large surveys beyond
$z \sim 3$? As always, the answer depends on what it is one wants to
learn. It is conceptually straightforward to {\it locate} higher
redshift galaxy candidates using variations on the Lyman break technique,
particularly with good sensitivity in the near-IR for redshifts beyond
$ z\sim 5$ or so. However, practical matters will probably prevent 
large and successful surveys useful for examining large scale structure.
We have recently completed a pilot spectroscopic survey for LBGs at
redshifts $z \simgt 4$ (Steidel \et 1999), and find that the surface density of candidates objects
to ${\rm I_{AB}} =25.0$ is just barely high enough to take advantage of
multiplexing using imaging spectrographs on 8--10m class telescopes.  Fainter
than this, because of the much brighter background one must fight
to get spectra of the higher redshift objects (the features that secure the
redshifts tend to be in the range $1200-1700$ \AA\ in the rest--frame) ,
anything close to spectroscopic completeness would be extremely
painful. On the other hand, if one can get around the more significant contamination
by interlopers in the $z \sim 4$ samples ($\sim 20\%$) it might be possible to
use de--projection of the angular correlation function of photometrically--selected
candidates to quite faint magnitudes
using the new generation of wide--field imagers. 
The question is whether the clustering of similarly--selected objects at $z \sim 4$ 
is providing much additional information over the much more easily--easured statistics
at $z \sim 3$. 

On the other hand, it has been possible to compare the properties of the galaxies
even in modest--sized samples at $z \sim 4$ and $z \sim 3$ using the ground--based
surveys. This is a bit of a diversion from the topic of large--scale structure in
general, but perhaps serves as an interesting example of how sample variance
can lead to somewhat misleading results, and in any case allows me to present
a new result that had just been obtained at the time of the meeting in August 1998. 
Most readers would be familiar
with the very exciting results from the Hubble Deep Field regarding the
star formation history of the universe as revealed by various galaxy surveys,
with the highest redshift points being obtained using Lyman break galaxies
within the $\sim 5$ square arc minute HDF. The implication from the work
of Madau \et (1996) and follow--up papers is that the UV luminosity density of
the universe, a proxy for the total SFR as a function of cosmic epoch, reached
a peak somewhere in the neighborhood of $z \sim 2$ and declined steadily beyond
that redshift. Fearing that perhaps the HDF might not be a representative region
of the universe, we have compiled a sample covering about 830 square arc minutes
($\sim 160$ times larger area than the HDF)
to the relatively bright magnitude of ${\rm I_{AB}}=25.0$, using photometric
selection designed to be as analogous as possible to the one implemented at
$z \sim 3$, and spectroscopic redshifts for about 50 $z \sim 4$ galaxies to
secure the redshift selection function. 
A comparison of the luminosity density represented in the bright
ends of the $z \sim 3$ and $z \sim 4$ luminosity functions indicates that
the luminosity density is essentially {\it constant} in the two redshift
ranges (see Figure 3), and indications are that the HDF is under--dense
in $z \sim 4$ galaxies relative to a survey covering a much larger volume. 
For amusement, I have reproduced a figure showing the latest incarnation of
the ``star formation history diagram'' in Figure 4, showing previous results
as well as the new points that come purely from the large ground--based
surveys. The moral of the story is that one can never survey too much
volume, and one should always be concerned about the lumpiness of the universe
when convincing oneself that one is seeing something ``universal''.

\section{Concluding Remarks}

Whereas the evolution of clustering of galaxies used to be seen
as a sure--fire cosmological test, this somewhat naive view 
has at this point probably gone the way of all other cosmological
tests that hoped to ignore the vagaries of galaxy evolution and
treat galaxies like test particles. If there are any points that
I would hope to get through in this highly qualitative talk
(the data do not yet justify anything much more sophisticated!),
it is that one needs to cover large volumes in order to hope to have
reliable estimates of even the simplest statistics, one needs to
worry a great deal about comparing apples to oranges in comparing
galaxy samples at different cosmic epochs, and one must remain
highly suspicious of what galaxy clustering is telling one about
the development of structure. It is essential to be able to isolate
cosmic epochs to avoid being overwhelmed by complicated projection
effects--- photometric redshifts seem like a very powerful tool
for surveying both large volumes and being able to slice a survey
in ways that will reveal what is really going on. Even better are
large spectroscopic surveys, in combination with photometric redshifts. 

In the end, because of complex epoch--dependent, luminosity
dependent, type--dependent bias, which is expected theoretically
(and now seen observationally) to be worse at high redshift,
one
must accept the fact that galaxy clustering may never constitute
a powerful cosmological test, even with all of the fantastic data
that will continue to roll in over the next several years. On the bright
side, it seems that our basic ideas about how galaxies form within halos
of dark matter whose distribution is easily understood using relatively
simple statistics or N-body simulations are holding up very well, and
(from an observer's point of view, at least) it is very encouraging
that theorists and observers seem by and large to be talking about the
same universe. There is enormous potential for progress in the area of understanding
the interface between galaxy formation and structure formation, and it
will involve a lot of interaction between theory and observations.  

Studying large scale structure by using galaxies ultimately
involves having to understand galaxies themselves, and the problems of structure formation
and galaxy formation are intimately related, and largely inseparable, problems. The very
obvious galaxy bias seen in the high redshift samples (at least, within the context
of generic models in which structure grows by gravitational instability from initial Gaussian perturbations) 
have perhaps emphasized the problem that has long been implicit in the theory--that galaxies
are not to be trusted as reliable tracers of mass, and you should trust the young ones
even less than the older ones. On the other hand, as eloquently pointed out by Carlos
Frenk in his closing remarks, it is human nature that one's interest in 
anyone (or anything) that one understands or trusts
quickly wanes, whereas the mysterious and untrustworthy seem all the more attractive, despite
one's better judgment.
This probably means that many cosmologists will be trying to understand galaxy formation after
all of the cosmological parameters are sorted out by MAP and Planck.  

\section*{Acknowledgments}

I'd like to thank the organizers for the invitation to attend this very interesting
meeting, and also for their patience in waiting for this wayward contribution.  
Thanks to Kurt Adelberger for invaluable comments on a draft of this article.

\section*{References}

\begin{thedemobiblio}{}
\japref Adelberger K., et al., 1998, ApJ, 505, 18
\japref Bagla J.S., 1998b, MNRAS, 299, 417
\japref Bagla J.S., 1998a, MNRAS, 297, 251
\japref Baugh, C.M., Cole, S., Frenk, C.S., and Lacey, C.G. 1998, ApJ, 498, 504
\japref Brainerd, T.G., Smail, I.S., and Mould, J.R. 1995, MNRAS, 275,781 
\japref Brainerd, T.G., and Villumsen, J.V. 1994, ApJ, 431, 477
\japref Carlberg R.G. et al., 1998, astro-ph/9805131
\japref Carlberg, R.G., Cowie, L.L., Songaila, A.m and Hu, E.M. 1997, ApJ, 483, 538
\japref Cohen, J.G., et al., ApJ, 1999,in press (astro-ph/9809067)
\japref Cole, S., Ellis, R.S., Broadhurst, T., Colless, M. 1994, MNRAS, 267, 541
\japref Coles, P., Lucchin, F., Matarrese, S., and Moscardini, L. 1998, MNRAS, 300, 183
\japref Connolly, A.J., et al., 1997, ApJ, 486, L11
\japref Davis, M., and Faber, S.M. 1998, in Wide Field Survyes in Cosmology, eds.
S. Colombi and Y. Mellier, in press (astro-ph/9810489)
\japref Efstathiou, G. 1995, MNRAS, 272, L25
\japref Efstathiou, G., et al., 1991, ApJ, 380, L47
\japref Eke, V.R., Cole, S., and Frenk, C.S. 1996, MNRAS, 282, 263
\japref Giavalisco, M. et al. 1999, ApJ, submitted
\japref Giavalisco, M., et al., 1998, ApJ, 503, 543
\japref Governato, F., et al. 1998, Nature, 392, 359 
\japref Groth, E.J., Peebles, P.J.E. 1977, ApJ, 217, 385
\japref Hogg, D. et al., 1998, AJ, 115, 1418
\japref Jing, Y.P., and Suto, Y. 1998, ApJ 494, L5
\japref Kaiser N., 1984, ApJ, {284}, L9
\japref Katz, N., Hernquist, L., and Weinberg, D.H. 1998, ApJ, submitted (astro-ph/9806257)
\japref Le F\`evre O., et al., 1996. ApJ, 461, 534
\japref Lilly, S.J., Le F\`evre, O., Hammer, F., and Crampton, D. 1996, ApJ, 460, L1
\japref Loveday, J., et al., ApJ, 442, 457
\japref Madau, P., Pozzetti, L., and Dickinson, M. 1998, ApJ, 498, 106 
\japref Madau, P., et al., 1996, MNRAS, 283, 1388
\japref Mo, H.J., Mao, S., and White, S.D.M. 1998, MNRAS, submitted (astro-ph/9807341)
\japref Mo H.J., Fukugita M., 1996, {\apj}, {467}, L9
\japref Mo H.J., White S.D.M., 1996, MNRAS, 282, 1096
\japref Oke, J.B. et al. 1995, PASP, 107, 375
\japref Peacock, J.A., and Dodds, S.J. 1996, MNRAS, 280, L19
\japref Pettini, M., et al., 1998, ApJ, in press (astro-ph/9806219)
\japref Postman, M., et al., 1998, ApJ, 506, 33
\japref Shepherd, C.W., et al., 1996 ApJ, 479, 82
\japref Steidel, C.C., et al., 1999, ApJ, submitted
\japref Steidel C.C., et al., 1998a, ApJ, 492, 428
\japref Steidel, C.C., et al., 1998b, Phil. Trans. R.S., in press (astro-ph/9805267)
\japref Steidel, C.C., et al., 1996, ApJ, 462, L17
\japref Steidel, C.C., Pettini, M., and Hamilton, D. 1995, AJ, 110, 2519
\japref Wechsler, R.H., et al., 1998, ApJ, 506, 19
\japref White S.D.M., Efstathiou G., Frenk C.S., 1993, {\mn}, {262}, 1023
\end{thedemobiblio}{}
\end{document}